# Spin-orbit torque magnetometry by wide-field magneto-optical Kerr effect

Tsung-Yu Tsai, Tian-Yue Chen, Chun-Ting Wu[†], Hsin-I Chan, and Chi-Feng Pai[*]

*Department of Materials Science and Engineering, National Taiwan University, Taipei 10617, Taiwan*

Magneto-optical Kerr effect (MOKE) is an efficient approach to probe surface magnetization in thin film samples. Here we present a wide-field MOKE technique that adopts a Köhler illumination scheme to characterize the current-induced damping-like spin-orbit torque (DL-SOT) in micron-sized and unpatterned magnetic heterostructures with perpendicular magnetic anisotropy. Through a current-induced hysteresis loop shift analysis, we quantify the DL-SOT efficiency of a Ta-based heterostructure with bar-shaped geometry, Hall-cross geometry, and unpatterned geometry to be $\left|\xi_{DL}\right| \approx 0.08$. The proposed wide-field MOKE approach therefore provides an instant and direct characterization of DL-SOT, without the need of any further interpretation on electrical signals.

---

[*] Email: cfpai@ntu.edu.tw
[†] Present address: Department of Computer Science and Information Engineering, National Taiwan University, Taipei 10617, Taiwan

Current-induced spin-transfer torque (STT) and spin-orbit torque (SOT) are efficient driving mechanisms for manipulating magnetization in various types of submicron-sized and micron-sized magnetic heterostructures[1-3]. Traditionally, to explore the magnetization dynamics and switching behaviors induced by these current-induced torques, electrical signals such as giant magnetoresistance and tunneling magnetoresistance are detected from patterned nano-devices like spin-valves and magnetic tunnel junctions[4-6]. More recently, various techniques were developed to characterize the SOT efficiency in micron-sized magnetic heterostructures, such as spin-torque ferromagnetic resonance[7], harmonic voltage measurement[8-10], spin-Hall magnetoresistance[11-13], unidirectional spin-Hall magnetoresistance[14]. However, quantification of SOT efficiencies from these techniques typically rely on further interpretations of the raw data and/or fittings with more than one parameter. A more direct characterization of SOT efficiency, particularly for heterostructures with perpendicular magnetic anisotropy (PMA), can be achieved by observing the dynamics of a chiral domain wall under the concurrent influence of applied charge current and in-plane magnetic field. This approach can be employed on either nano-wires (spin hall torque magnetometry)[15] or micron-sized Hall-cross devices (hysteresis loop shift measurement)[16], and has been used by several different groups to characterize SOT efficiencies in transition metal/ferromagnetic metal[17], transition metal/ferrimagnetic metal[18] as well as topological insulator/ferrimagnetic metal[19] bilayer systems.

On the other hand, probing SOT or its corresponding effective field through optical means can serve as an alternative for SOT efficiency quantification that is virtually free from the artifacts in electrical signals. It has been shown that SOT magnetometery can be realized by detecting the SOT-induced magnetization dynamics through magneto-optical Kerr effect (MOKE) in either harmonic mode[20-22] or pump-probe mode[23] with lasers. In this work, we develop an optical SOT magnetometry protocol, which utilizes the advantage of Köhler illumination and wide-field polar MOKE to characterize the SOT efficiencies of micron-sized bar-shaped samples and unpatterned samples under an optical microscope. By measuring samples with a two-terminal bar-shaped geometry, we further quantify the damping-like SOT (DL-SOT) efficiency of a Ta-based magnetic heterostructure to be $|\xi_{DL}| = 0.077 \pm 0.003$, which is greater than the electrically determined $|\xi_{DL}| \approx 0.04$ in one of our previous studies using Hall-cross devices[24]. By probing the very same four-terminal Hall-cross device both electrically and optically, we find that this discrepancy is caused by the shunting of current into the voltage arms of Hall-cross devices. If we take the shunting effect into account, a more consistent value of $|\xi_{DL}| \approx 0.08$ can be obtained from both electrical and optical approaches. More importantly, this optical SOT magnetometry can be further applied to unpatterned (as-prepared) thin films, which has never been realized before. The versatility of this optical SOT characterization protocol therefore provides us a simple way to quantify DL-SOT efficiency free from parasitic electrical signals that might undermine a more accurate estimation.

Our major experimental setup consists of an optical microscope with LED (white light)

illumination of Köhler configuration, as shown in Fig. 1(a). The magnification of the microscope is set to be 100x and the working distance is around 1.5 cm. Two polarizers are separately installed in front of the LED light source and in front of the CCD camera, which are respectively called polarizer and analyzer. In this scheme, the white light from LED is polarized by the polarizer, then shine on the sample of interests. The polarized light then interacts with the magnetization in sample and reflects back to the microscope. After the reflected light passes through the analyzer, it is captured by the CCD. In short, this is a polar MOKE measurement setup with additional optical microscope functions. If dealing with magnetic materials with PMA, then the CCD should pick up an intensity variation when the magnetization switches from pointing up-ward to pointing down-ward. The program we have developed allows us to obtain the intensity from CCD (as the sum of red, green, and blue values from pixels on screen) while sweeping an external out-of-plane magnetic field $H_z$[25]. An exemplary hysteresis loop raw data acquired from a sputter-deposited Ta(4)/Co$_{20}$Fe$_{60}$B$_{20}$(1.4)/Hf(0.5)/MgO(2) PMA film (annealed at 300°C for 1 hour, numbers in parenthesis are in nm) by this protocol is shown in Fig. 1(b). Note that the Hf 0.5 nm insertion layer is employed to enhance the PMA[26]. Since both the averaged detected intensity $I$ and the intensity change due to magnetization orientation $\Delta I$ should vary with respect to the angle between polarizer and analyzer $\theta$, the figure of merit for detected MOKE signal, $\Delta I / I = 2K_r \sin 2\theta / (\cos^2 \theta + \gamma)$ [27]. $K_r$ stands for the material dependent Kerr rotation and $\gamma$ represents the depolarization fraction due to the imperfection of polarizers. For our measurement setup, the largest $\Delta I/I \approx 0.02$ can be achieved at $\theta \approx 80$ degrees and the angle dependence of $\Delta I/I$ can be well-fitted to $\Delta I/I = 2K_r \sin 2\theta/(\cos^2\theta + \gamma)$ with $K_r = 8\times 10^{-4}$ rad (45.8 mdeg) and $\gamma = 4.2\times 10^{-3}$ for this particular sample, as shown in Fig. 1(c). $\theta$ is therefore fixed at around 80 degrees for the best signal-to-noise ratio for the rest of this work.

In order to characterize DL-SOT efficiencies of magnetic heterostructures optically, we first turn our focus on patterned micron-sized devices. As schematically shown in Fig. 2(a), the above-mentioned Ta/CoFeB/Hf/MgO film is patterned into bar-like devices with lateral dimensions of $10\,\mu\text{m}\times 20\,\mu\text{m}$ using photolithography and lift-off processes. Ta(4)/Cu(100) are deposited as contact pads for electrical probes. A representative view of a $10\,\mu\text{m}\times 20\,\mu\text{m}$ bar-like device under optical microscope is shown in Fig. 2(b). Using our program, we can select an arbitrary area-of-interest on screen for signal acquisition, which is referred as the probing area. We use two tungsten probes to send electrical/charge current into the bar-like device, and pick up the magnetization response through wide-field MOKE signal (intensity) within the probing area.

Based on the concept of current-induced hysteresis loop shift measurement[16], the DL-SOT efficiency of a micron-sized PMA heterostructure can be estimated from the shift of out-of-plane hysteresis loop $H_z^{\text{eff}}$, which is traditionally detected via anomalous Hall voltage, with respect to the applied charge current $I_{dc}$. Similar to the all-electrical detection approach, the applied $I_{dc}$ can also generate Joule heating and possibly leads to the reduction of coercive field ($\Delta H_c \propto I_{dc}^2$). However, if we only consider the shift of the center of hysteresis loop ($H_z^{\text{eff}} \propto I_{dc}$), then the heating effect can be eliminated and only the DL-SOT effect will be seen. Typically an in-plane field $H_x$ greater than the interfacial Dzyaloshinskii-Moriya interaction (DMI) effective field $H_{DMI}$ is required to realign the chiral domain walls in the magnetic layer[15,28], such that the saturated DL-SOT efficiency $|\xi_{DL}| \propto |H_z^{eff} / I_{dc}|$ can be achieved. In the wide-field MOKE approach, we obtain out-of-plane hysteresis loop shifts via optical intensity signals, instead of Hall voltages, while applying charge currents across the bar-like sample. An in-plane field $H_x > H_{DMI} \approx 100\,\text{Oe}$ is applied to see the full effect from DL-SOT. Two MOKE-obtained representative out-of-plane hysteresis loops under positive and negative charge currents with $H_x = 600\,\text{Oe}$ are shown in Fig. 2(c), for a $10\,\mu\text{m}$-wide bar-like device. Note that the applied in-plane field is smaller than the in-plane saturation field $H_{sat} \approx 1300\,\text{Oe}$ for this particular PMA sample. By summarizing $H_z^{eff}$'s as functions of $I_{dc}$, as shown in Fig. 2(d), linear trends of opposite slopes for $H_x = \pm 600\,\text{Oe}$ can be clearly seen. This is consistent with the results from a Hall voltage approach[16]. The opposite slopes are caused by the opposite polarities of realigned domain walls.

To quantify DL-SOT efficiency of this particular Ta/CoFeB bilayer sample from the fitted averaged slope of $H_z^{eff} / I_{dc} = -(6.97 \pm 0.26)\,\text{Oe/mA}$, we use[16,29]

$$\xi_{DL} = \frac{2e}{\hbar}\left(\frac{2}{\pi}\right)\mu_0 M_s t_{CoFeB}^{eff} w t_{Ta}\left(\frac{\rho_{CoFeB} t_{Ta} + \rho_{Ta} t_{CoFeB}}{\rho_{CoFeB} t_{Ta}}\right)\left(\frac{H_z^{eff}}{I_{dc}}\right), \quad (1)$$

where $M_s \approx 1.28 \times 10^6 \text{ A/m}$, $t_{\text{CoFeB}}^{\text{eff}} \approx 0.83 \text{ nm}$, $\rho_{\text{CoFeB}} \approx 200 \mu\Omega \cdot \text{cm}$ [24] respectively stand for saturation magnetization, effective thickness (nominal thickness subtracted by the thickness of magnetic dead layer of 0.57 nm), and resistivity of the ferromagnetic CoFeB layer. The magnetic properties of the heterostructure are characterized by vibrating sample magnetometery from a series of Ta(4)/CoFeB($t_{\text{CoFeB}}$)/Hf(0.5)/MgO(2) films[24]. $t_{\text{Ta}} = 4 \text{ nm}$ and $\rho_{\text{Ta}} \approx 200 \mu\Omega \cdot \text{cm}$ represent thickness and resistivity of the Ta spin-Hall metal layer, respectively. $w = 10 \mu\text{m}$ is the width of the bar-shaped device. Note that we neglect the contribution from Hf(0.5) dusting layer since it should be discontinuous and hardly contributes to the charge and/or spin transport of the whole heterostructure. The DL-SOT efficiency of this Ta-based heterostructure is therefore estimated to be $\xi_{DL} = -0.077 \pm 0.003$. From our measurements on various bar-like devices with different widths (ranges from $5 \mu\text{m}$ to $10 \mu\text{m}$), the DL-SOT efficiency values extracted by this appraoch does not depend on the size of the area-of-interest. We also employ the same procedure on a series of W-based heterostructure with PMA (W/CoFeB/Hf/MgO) and find $\xi_{DL} = -0.158 \pm 0.005$, which is greater in magnitude while comparing to that of Ta. In addition to efficiency characterization, we also demonstrate that current-induced SOT switching of magnetization can also be observed optically, as shown in Fig. 3(a) and (b), for $H_x = \pm 100 \text{ Oe}$. The symmetries of the switching curves are consistent with those being obtained electrically from a similar Ta-based heterostructure[2], which again verifies the feasibility of studying SOT-related phenomenon via wide-field optical means. Note that the magnitude of estimated DL-SOT efficiency for Ta-based heterostructures, $|\xi_{DL}| \approx 0.08$, is greater than one of our previously reported values of $|\xi_{DL}| \approx 0.04$, in which the magnetization response was obtained electrically through a Hall-bar device[24]. We believe that this is caused by the difference in geometry of the devices employed: For a Hall-bar or Hall-cross device, the applied current flowing across the intersecting region might not be entirely confined by the nominal width $w$ of the bar, therefore causing an overestimation on the current density. That is, the effective width near the intersection of a Hall-bar device $w_{eff} > w$ and $0.04$ only represents the lower bound of the real $|\xi_{DL}|$. In contrast, the two terminal bar-shaped device for wide-field MOKE detection is free from this geometrical factor, therefore resulting in a larger and perhaps a more accurate estimation on DL-SOT efficiency.

To further verify the existence of possible current shunting in Hall-cross devices, we perform both electrical (anomalous Hall effect, AHE) and optical (MOKE) hysteresis loop shift measurements on the very same micron-sized Ta/CoFeB/Hf/MgO Hall-cross device with a nominal width of $w = 5\ \mu m$. As shown in Fig. 4(a) and (d), the electrical detection of loop shift results in $H_z^{eff}/I_{dc} = -(8.96 \pm 0.21)\ \text{Oe/mA}$. If we assume the current flow width is the same as the nominal width of Hall-cross, $w = 5\ \mu m$, then the estimated DL-SOT efficiency based on Eqn. (1) is $\xi_{DL} = -0.049 \pm 0.003$. A similar number of $\xi_{DL} = -0.053 \pm 0.002$ can be obtained by probing the hysteresis loop shift optically (wide-field MOKE signal) around the intersection of the Hall-cross device, as shown in Fig. 4(b) and (e). However, if we set our probing area-of-interest to the "bar region" (Fig. 4(c) and (f)), then the estimated magnitude of DL-SOT efficiency becomes larger, $\xi_{DL} = -0.084 \pm 0.002$. These observations are consistent with what we speculated: The effective width of current flow $w_{eff}$ around the Hall-cross intersection region is indeed greater than the nominal width $w = 5\,\mu m$ due to possible shunting of currents into the voltage arms of Hall-cross devices, therefore results in a smaller apparent estimated $|\xi_{DL}|$. If we further assume that the actual DL-SOT efficiency $\xi_{DL}$ is uniform across the sample, then the effective width of current flow around the Hall-cross intersection is estimated to be $w_{eff} \approx w \cdot (0.084/0.053) \approx 1.58w$. In this case, $w_{eff} \approx 7.9\,\mu m$.

Next, in order to show possible advantages of using wide-field MOKE DL-SOT magnetometry, we execute the same measurement protocol on unpatterned thin films. As shown in Fig. 5(a), two tungsten probes are brought into proximity and land on an unpatterned Ta/CoFeB/Hf/MgO heterostructure film with a separation of $D \approx 100\,\mu m$. The width and the length of the wide-field MOKE probing area are set to $w \approx 10\,\mu m$ and $L \approx 70\,\mu m$, respectively. The same procedure is carried out by stepping the applied current $I_{dc}$ from 8 mA to -8 mA, with the in-plane field fixed at $H_x = 500\,Oe$. If we collect the data from a probing area away from the region of stronger current flow (in between two probes), then no obvious current-induced shift can be observed, as presented in Fig. 5(b). In contrast, as shown in Fig. 5(c) and (d), wide-field MOKE successfully retrieve similar current-induced hysteresis loop shifts that resemble the trends observe in micron-sized devices within the region of stronger current flow, with $H_z^{eff}/I_{dc} = -(0.5 \pm 0.06)$ Oe/mA. To further determine DL-SOT efficiency $|\xi_{DL}|$ from the measured $H_z^{eff}/I_{dc}$ with Eqn. (1), we will also need a good estimation on the effective width $w_{eff}$ of current flow in this type of two-probe configuration.

Since the typical distance between two probe tips $D$ is much greater than the total thickness of the conductive layer of deposited film $t = t_{CoFeB} + t_{Ta} \approx 5.4$ nm, it is reasonable to assume that the current will be flowing in a 2-dimensional manner. The electric field $\vec{E}(r)$ of an arbitrary point on the thin film with respect to the sourcing probe can be written as $\vec{E}(r) = (I_{dc}\rho/2\pi rt)\hat{r}$, where $I_{dc}$, $\rho$, and $r$ represent the applied DC current, effective resistivity of the film ($\rho \approx \rho_{CoFeB} \approx \rho_{Ta} \approx 200\,\mu\Omega\cdot$cm), and the distance between sourcing probe and point-of-interest, respectively. If we consider both sourcing and draining probes and choose the midpoint between them to be origin, as shown in Fig. 6(a), the electric potential of an arbitrary point on the film can be expressed as $V(x,y) = (I\rho/2\pi t)\ln\left[\sqrt{(x-D/2)^2+y^2}/\sqrt{(x+D/2)^2+y^2}\right]$ and the vector field of current density on the film can be calculated by $\vec{J}(x,y) = -\vec{\nabla}V(x,y)/\rho$. The most relevant component of $\vec{J}$ for our concern is along $x$ direction, and can be expressed as $J(x,y) = J_x(x,y) = -\rho^{-1}\partial V(x,y)/\partial x$ (See Fig. 6(b)). The average current density along $x$ direction within the probing area-of-interest then can be expressed as

$$\langle J \rangle = -\frac{I}{2\pi twL}\int_{-w/2}^{w/2}\int_{-L/2}^{L/2}\frac{\partial}{\partial x}\ln\left[\sqrt{(x-D/2)^2+y^2}/\sqrt{(x+D/2)^2+y^2}\right]dxdy, \qquad (2)$$

which can be simplified to

$$\langle J \rangle = \frac{I}{2\pi twL} \int_{-w/2}^{w/2} \ln\left[\frac{\left(\frac{L+D}{2}\right)^2 + y^2}{\left(\frac{L-D}{2}\right)^2 + y^2}\right] dy. \tag{3}$$

Therefore, the effective width of the current flow for this two-probe configuration is

$$w_{\text{eff}}(w, L, D) \equiv \frac{I}{t\langle J \rangle} = 2\pi wL \left\{\int_{-w/2}^{w/2} \ln\left[\frac{\left(\frac{L+D}{2}\right)^2 + y^2}{\left(\frac{L-D}{2}\right)^2 + y^2}\right] dy\right\}^{-1}. \tag{4}$$

Using $D \approx 100\,\mu\text{m}$, $w \approx 10\,\mu\text{m}$, and $L \approx 70\,\mu\text{m}$, the effective width is determined to be $w_{\text{eff}} \approx 128\,\mu\text{m}$. The DL-SOT efficiency is further estimated from Eqn. (1) with this $w_{\text{eff}}$ and the measured $H_z^{\text{eff}}/I_{dc} = -(0.5 \pm 0.06)$ Oe/mA to be $|\xi_{DL}| \approx 0.071$. We also perform measurements on the same film with different $D$ and $L$ ($\approx D - 30\,\mu\text{m}$) with the probing zone width fixed at $w \approx 10\,\mu\text{m}$. By changing the distance between two probes $D$ as well as the probing zone length $L$, we observe a variation of measured $H_z^{\text{eff}}/I_{dc}$, as shown in Fig. 6(c). This decreasing trend of $H_z^{\text{eff}}/I_{dc}$ as a function of $L$ can be predicted from Eqn. (1) with $|\xi_{DL}| = 0.074$ and the corresponding $w_{\text{eff}}$ ($w = 10$ μm, $5$ μm $\leq L \leq 70$ μm, $35$ μm $\leq D \leq 100$ μm). If we directly estimate $|\xi_{DL}|$ from the calculated $w_{\text{eff}}$ and the measured $H_z^{\text{eff}}/I_{dc}$, then the overall DL-SOT efficiency is determined to be $|\xi_{DL}| = 0.074 \pm 0.015$ based on Eqn. (1), as shown in Fig. 6(d). This value is fairly close to what we obtained from the patterned bar-shaped device, $|\xi_{DL}| = 0.077 \pm 0.003$. A more rigorous calculation on the current distribution and the resulting $w_{\text{eff}}$, for instance, by further taking the probe tip size effect into account, will perhaps lead to an even more accurate estimation on $|\xi_{DL}|$ for the unpatterned systems. Also note that for patterned micron-sized samples, imperfections such as device edge roughness and curvature might influence the accuracy of DL-SOT efficiency estimation. These issues could be possibly solved by using unpatterned films for DL-SOT characterization.

In summary, we show that using wide-field polar MOKE with Köhler configuration is a simple and efficient approach for SOT characterization, especially in quantifying the DL-SOT efficiency $|\xi_{DL}|$ from micron-sized bar-shaped magnetic heterostructures with PMA. This allows us to probe and study SOT-driven dynamics from devices with simple geometries, and the obtained data will be free from electrical artifacts. Also note that although our focus in the present work is DL-SOT characterization, technically, the MOKE detection scheme can be extended to characterize field-like SOT. One possible approach will be measuring the current-induced in-plane hysteresis loop for heterostructures with PMA. The protocol of wide-field MOKE DL-SOT magnetometry can be further adopted for unpatterned thin films, and the estimated DL-SOT efficiencies are fairly close to those obtained from patterned devices. Our results obtained from samples with bar-shaped ( $|\xi_{DL}| = 0.077 \pm 0.003$ ), Hall-cross ( $|\xi_{DL}| = 0.084 \pm 0.002$ ), and unpatterned ( $|\xi_{DL}| = 0.074 \pm 0.015$ ) geometries indicate that the DL-SOT efficiency of a Ta/CoFeB/Hf/MgO heterostructure is $|\xi_{DL}| \approx 0.08$. Further improvements on this technique could possibly lead to an even more accurate fast-track DL-SOT characterization on as-prepared magnetic heterostructures without any further device fabrication processes.

**Data Availability**

The datasets generated during and/or analysed during the present study are available from the corresponding author on reasonable request.

**Acknowledgments**

This work is supported by the Ministry of Science and Technology of Taiwan (MOST) under Grant No. MOST 105-2112-M-002-007-MY3.

**Author contributions**

C.-F. P. proposed, designed, and supervised the study. T.-Y. T. and T.-Y. C. performed the measurements. T.-Y. C. prepared all the tested samples. T.-Y. T., C.-T. W., and H.-I. C. built the measurement setup and related controlling programs. C.-F. P. and T.-Y. T. prepared the manuscript and figures. All authors reviewed the manuscript.

**Competing interests**

The authors declare no competing interests.

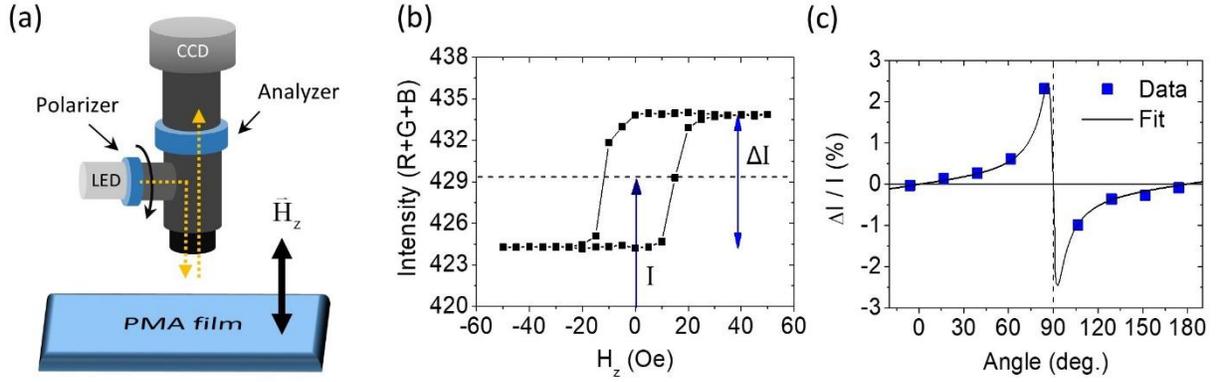

Figure 1. (a) Schematic illustration of the experimental setup for wide-field polar MOKE measurement with a Köhler illumination scheme (b) Exemplary out-of-plane hysteresis loop from a Ta(4)/CoFeB(1.4)/Hf(0.5)/MgO(2) unpatterned film. *I* represents the mean intensity whereas $\Delta I$ is the difference between maximum and minimum MOKE responses. (c) Kerr rotation of Ta(4)/CoFeB(1.4)/Hf(0.5)/MgO(2) as a function of the angle between polarization directions of polarizer and analyzer. The solid line is the fit of data to $2K_r \sin 2\theta / (\cos^2 \theta + \gamma)$ with $K_r = 8 \times 10^{-4}$ rad and $\gamma = 4.2 \times 10^{-3}$.

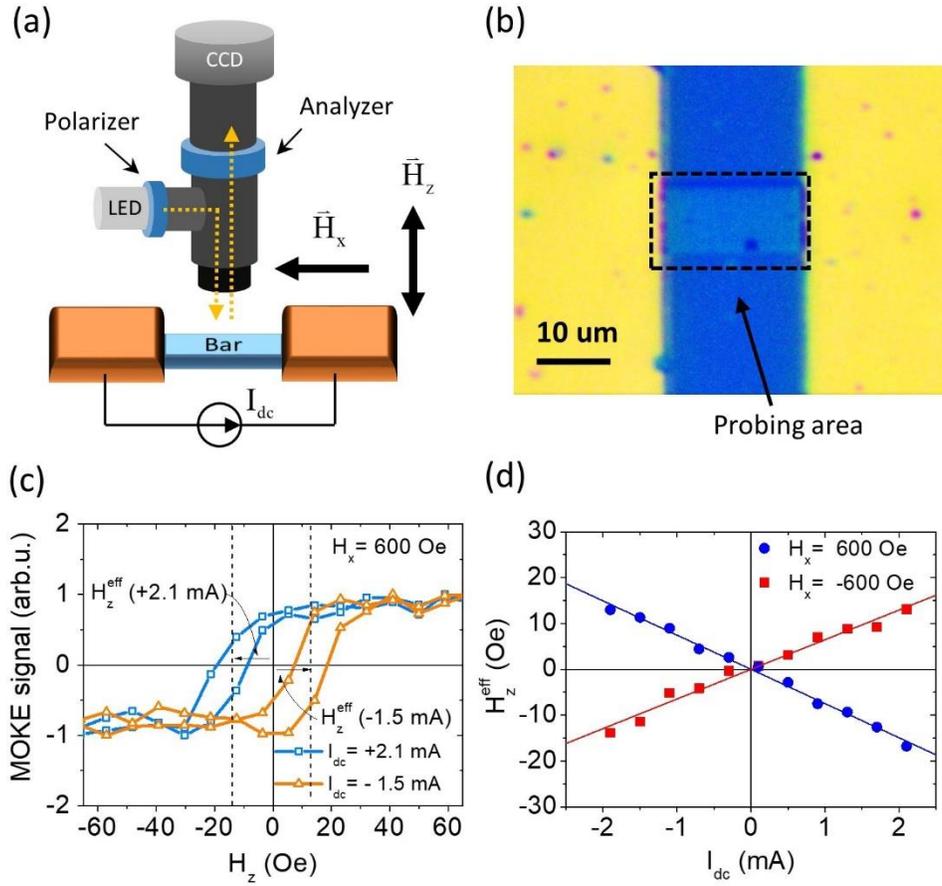

Figure 2. (a) Schematics of wide-field polar MOKE signal detection for a micron-sized bar-shaped sample under current-induced hysteresis loop shift measurements. (b) A representative image of the measured bar-shaped device ($10\,\mu m \times 20\,\mu m$) and the probing area under optical microscope. (c) Optically-detected hysteresis loops of a Ta(4)/CoFeB(1.4)/Hf(0.5)/MgO(2) sample with dc currents ($I_{dc}$) of opposite polarities and under the application of an in-plane bias field $H_x = 600$ Oe to overcome Dzyaloshinskii-Moriya interaction. $H_z^{eff}$'s represent the amount of shifts in terms of effective field due to current-induced SOT. (d) $H_z^{eff}$ as functions of $I_{dc}$ for $H_x = \pm 600$ Oe.

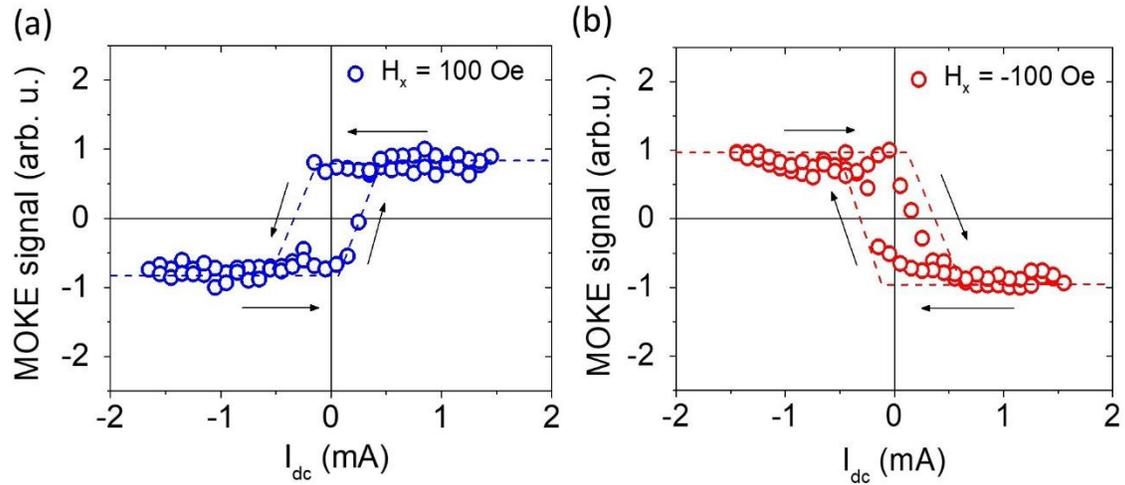

Figure 3. Exemplary wide-field MOKE-detected current-induced SOT switching curves of a Ta(4)/CoFeB(1.4)/Hf(0.5)/MgO(2) bar-shaped sample under the application of (a) $H_x = 100\,\text{Oe}$ and (b) $H_x = -100\,\text{Oe}$. The dashed lines serve as guides to the eye.

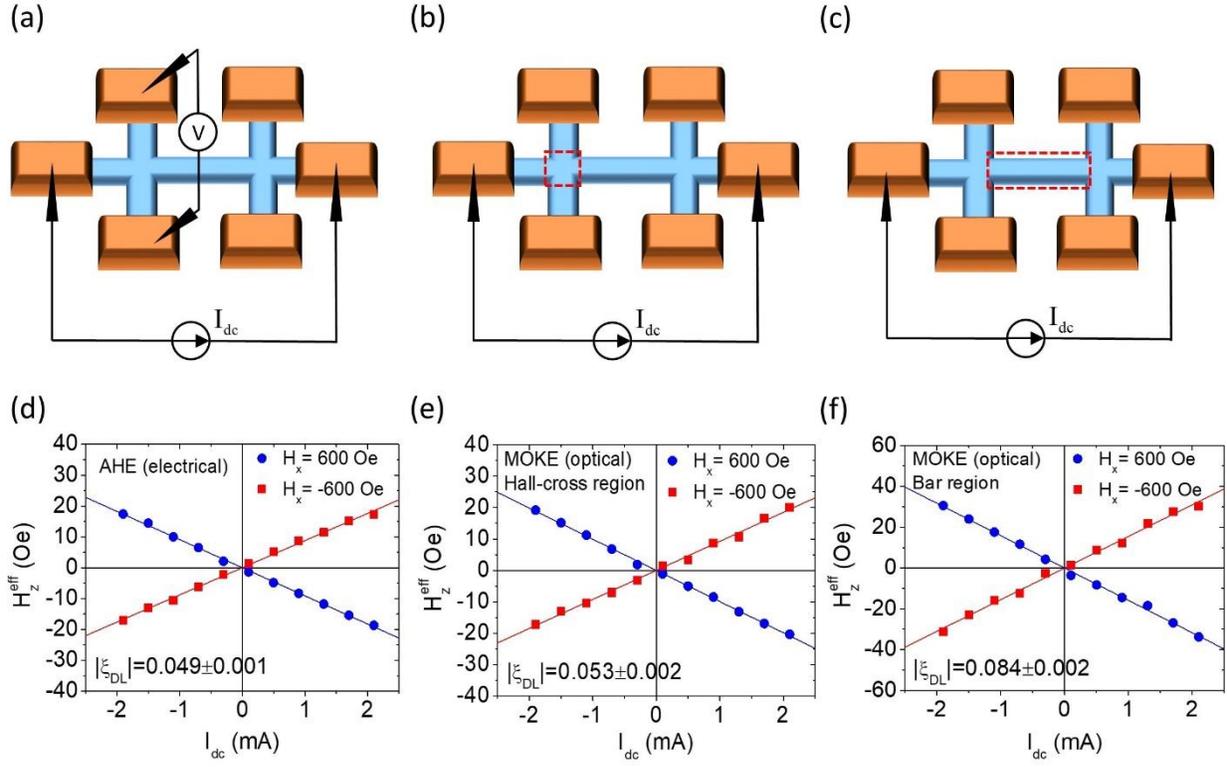

Figure 4. Schematic illustrations of DL-SOT effective field measurements through (a) electrical and (b, c) optical approaches on a Ta(4)/CoFeB(1.4)/Hf(0.5)/MgO(2) Hall-cross device. The red dashed boxes in (b) and (c) represent the MOKE detection areas-of-interest, namely the Hall-cross region and the bar region, respectively. (d, e, f) $H_z^{\text{eff}}$ as functions of $I_{dc}$ for $H_x = \pm 600$ Oe under different measurement schemes.

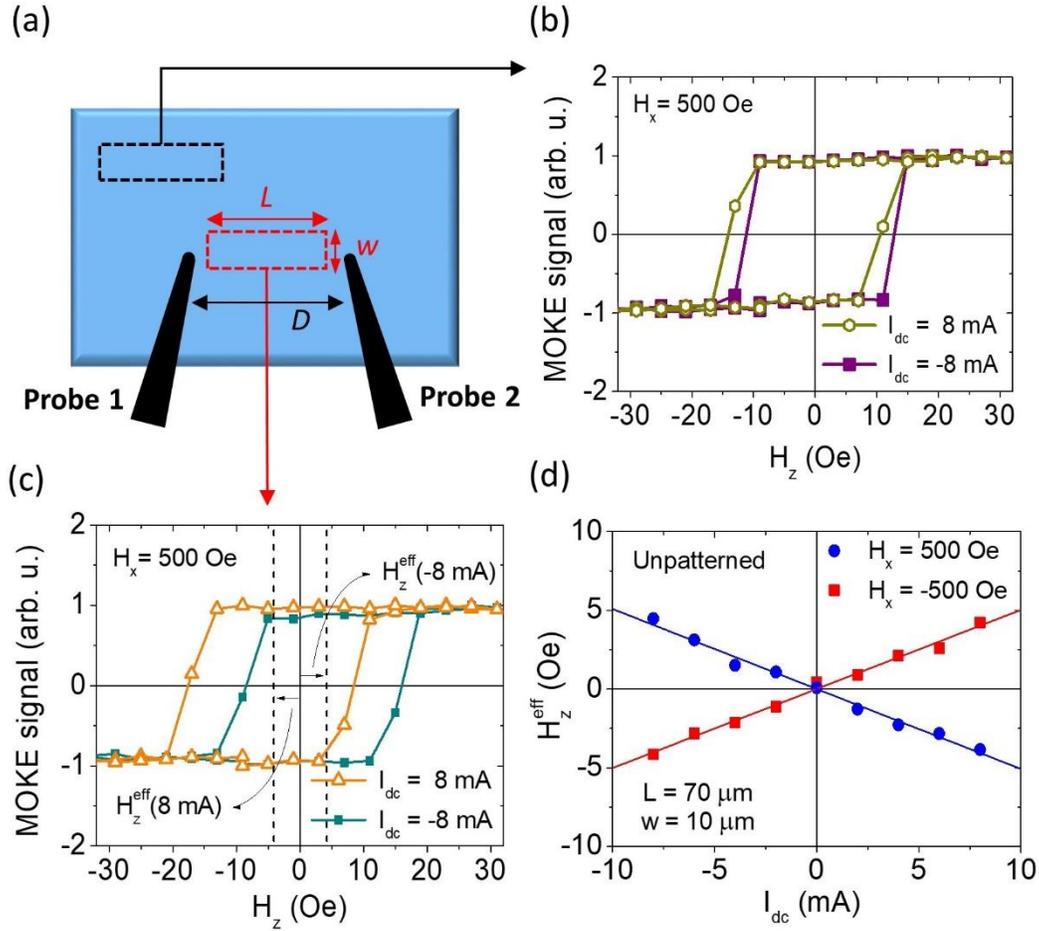

Figure 5. (a) Schematics (top view) of tungsten probes configuration for measuring SOT effective field in a Ta(4)/CoFeB(1.4)/Hf(0.5)/MgO(2) unpatterned film. The two dashed boxes represent two different probing areas. $w$ and $L$ represent the width and the length of the probing area, respectively. $D$ is the distance between two probe tips. Optically-detected hysteresis loops for $I_{dc} = \pm 8$ mA and $H_x = 500$ Oe from the probing area (b) away from the region with strong current flow and (c) within the strong current flow region. (d) $H_z^{eff}$ of the unpatterned Ta-based heterostructure as functions of $I_{dc}$ for $H_x = \pm 500$ Oe.

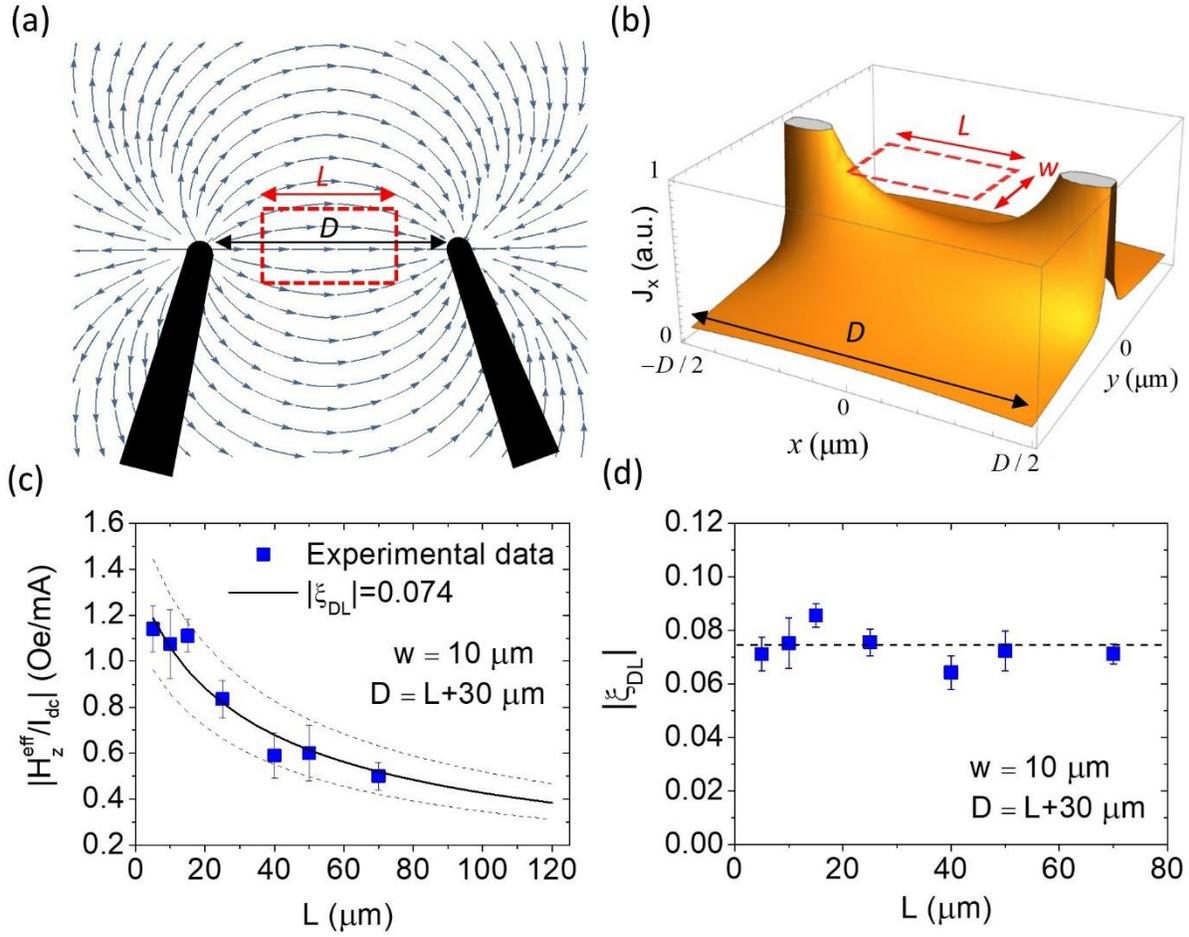

Figure 6. (a) Streamline plot of the current density vector field $\bar{J}(x,y)$ and (b) the calculated $x$ component of current density $J_x(x,y)$ for two-probe configuration. The dashed box represents the probing area. (c) Optically-determined $|H_z^{\text{eff}}/I_{dc}|$ plotted as a function of probing area length $L$ for a Ta(4)/CoFeB(1.4)/Hf(0.5)/MgO(2) unpatterned film. The solid line represents the predicted $|H_z^{\text{eff}}/I_{dc}|$ based on Eqn. (1) and the $L$-dependent effective width $w_{\text{eff}}$ with $|\xi_{DL}|=0.074$. The dashed lines represent the predicted trend lines for $|\xi_{DL}|=0.06$ and $|\xi_{DL}|=0.09$. (d) Estimated $|\xi_{DL}|$ as a function of $L$. The dashed line represents the mean of all results, $|\xi_{DL}|=0.074$.